\documentclass[12pt,a4paper]{article}

\usepackage[dvips]{graphics}

\begin{document}

\title{Superluminal optical pulse propagation in nonlinear coherent media }
\author{Ruben~G.~Ghulghazaryan$^1$ and Yuri~P.~Malakyan$^{2,}$\thanks{E-mail: yumal@ipr.sci.am} \\
{\small \textsl{$^1$ Department of Theoretical Physics, Yerevan Physics Institute,}} \\
{\small \textsl{Alikhanian Brothers 2, 375036 Yerevan, Armenia}}\\
{\small \textsl{$^2$ Institute for Physics Research, Armenian National Academy of Sciences,}} \\
{\small \textsl{ Ashtarak-2, 378410, Armenia}}}
\date{}
\maketitle
 
\begin{abstract}
The propagation of light-pulse with negative group-velocity in a nonlinear medium is studied theoretically. 
We show that the necessary conditions for these effects to be observable are realized in a three-level 
$\Lambda$-system interacting with a linearly polarized laser beam in the presence of a static magnetic field. 
In low power regime, when all other nonlinear processes are negligible, the light-induced Zeeman
coherence cancels the resonant absorption of the medium almost completely, but preserves the dispersion 
anomalous and very high. As a result, a superluminal light pulse propagation can be observed in the sense 
that the peak of the transmitted pulse exits the medium before the peak of the incident pulse enters. 
There is no violation of causality and energy conservation. Moreover, the superluminal effects are 
prominently manifested in the reshaping of pulse, which is caused by the intensity-dependent pulse velocity. 
Unlike the shock wave formation in a nonlinear medium with normal dispersion, here, the self-steepening 
of the pulse trailing edge takes place due to the fact that the more intense parts of the pulse travel 
slower. The predicted effect can be easily observed in the well known schemes employed for studying of 
nonlinear magneto-optical rotation. The upper bound of sample length is found from the criterion that 
the pulse self-steepening and group-advance time are observable without pulse distortion caused by 
the group-velocity dispersion.
    
PACS numbers: 32.80.Bx, 42.50.Gy, 42.50.Gt, and 42.25.Bs

\end{abstract}

\section{Introduction}

The propagation of light pulses through the media with a normal dispersion is adequately 
described by the group velocity, which in this case turns out to be, in fact, the velocity of signal 
transfer. However, the group velocity loses its role of signal velocity in the region of anomalous
dispersion, where it not only exceeds the vacuum speed of light, but at certain frequencies 
becomes negative, whereas, even when the group velocity is superluminal, no real signal can be
transferred faster than the vacuum velocity of light~\cite{Brillouin}. Nevertheless, contrary 
to the common assertion~\cite{Born,Jackson} that superluminal group velocity is not a useful concept,
Garrett and McCumber~\cite{Garrett} have shown that in the case of superluminal propagation 
of sufficiently smooth pulses, such as Gaussian wave packets,the motion of the pulse
peak is correctly described by the classical expression of group velocity. The effect of 
superluminality is that the emerging pulse has essentially the same shape and width 
as that of the incident wave packet, but its peak exits the cell before the incident pulse
even enters. This process can be understood in terms of superposition and interference of 
traveling plane waves that have a distribution of frequencies and add up to form a 
narrow-band light pulse. In a dispersive medium the speed of each plane wave is determined by the
refractive index at that frequency. Around the resonance in the region of anomalous dispersion
the refractive index varies so steeply and different plane waves interfere in a way
that the point of maximal constructive interference is localized at the exit face of the sample.
The plane waves appear in the medium much earlier and, hence, the exit pulse is formed long before
than the peak of incident pulse enters the medium. When the group velocity is negative the exit pulse
splits into two pulses, one of which moves forward in the vacuum, while the second one 
propagates backward in the medium and is canceled by the incoming pulse at the entrance of the cell 
(see Sec. 3).The superluminal propagation can be phenomenologically represented as reshaping of 
the pulse, most of which is absorbed, leaving only a small pulse in the leading edge. Eventually, 
the outgoing pulse moves superluminally and saves the shape of initial pulse. This picture breaks 
down, if the envelope of incident wave packet has singularities, for example, step-type modulations, 
that always moves with the light velocity $c$. All these results demonstrate that for smooth 
pulses with continuous derivatives the group velocity is physically meaningful even when it 
is negative. But it cannot be interpreted as a velocity of real signal, which could propagate 
no faster than light~\cite{Diener}. These predictions have been confirmed experimentally using 
various schemes, including the anomalous dispersion near an absorption line~\cite{Chu,Segard},
nonlinear gain~\cite{Fisher}, active plasma medium~\cite{Basov} and the tunneling of one-photon 
wave packet through a barrier~\cite{Steinberg}.

An interesting model for a transparent anomalous dispersive medium and a light pulse propagation
with negative group velocity has been proposed by Chiao and co-workers~\cite{Steinberg1, Bolda}. 
The model is based on the destructive interference between the two close-lying gain transitions. 
In this scheme L.Wang and co-workers have recently observed a pulse-advance shift of $62ns$ in 
cesium vapor~\cite{Dogariu}.

The results of papers~\cite{Garrett}-\cite{Fisher},~\cite{Steinberg}-\cite{Chiao} are based on 
linear response of the medium where the negative group 
velocity always exists within the absorption lines~\cite{Brillouin} and outside gain lines
~\cite{Steinberg1}-\cite{Chiao}. The similar results have been obtained, when a weak probe beam 
propagates with an effective index of refraction induced by a nonlinear interaction of the medium 
with a strong pump field. Under these conditions both the probe light ultraslow propagation
~\cite{Hau}-\cite{Budker} at the electromagnetically induced transparency (EIT)~\cite{Harris} 
and the negative "group-delay" propagation associated with electromagnetically induced opacity
~\cite{Akulshin, Lezama} have been recently observed experimentally. However, from the point of 
view of the dependence of optical properties of medium on probe field intensity these systems 
may be clearly considered as effectively linear optical media with an extraordinary large dispersion.

\begin{figure}[p]
\center
\includegraphics{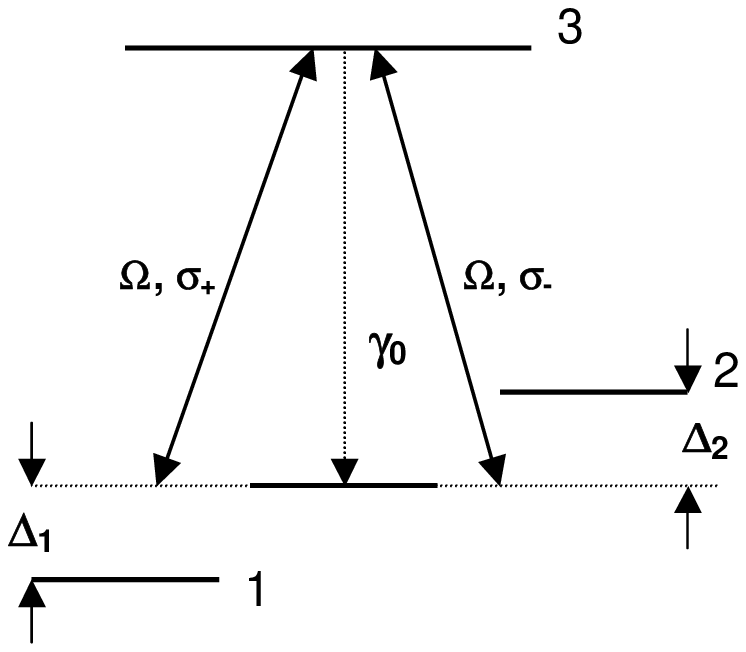}
\caption{Open $\Lambda$-configuration for studying the nonlinear superluminal light propagation. 
$\sigma _{\pm}$ are components of a linearly polarized laser field with Rabi frequency $\Omega$ coupling the 
level $|3\rangle$ with $|1\rangle$ and $|2\rangle$, respectively. The Zeeman shift of ground-state 
sublevels $\Delta_{1,2}=\mp\Delta$ is induced by the longitudinal magnetic field $B$. Radiative decay 
from $|3\rangle$ to $|1\rangle$ and $|2\rangle$ goes at rate $\gamma$ and the outside with the 
rate $\gamma_{0}$.
\label{fig1}}
\end{figure}

Intriguing questions arise whether the nonlinear superluminality occurs in a medium, the refractive 
index of which is modified by the propagating light pulse. How this effect is pronounced, and what is the 
form of superluminal pulse emerging from such a nonlinear medium? It seemed, that the answer to these 
questions could be easily obtained in a two-level atom, which is a basic model for the study of 
superluminal propagation in linear media~\cite{Garrett}-\cite{Fisher},\cite{Steinberg}. Notice that 
this system has been also employed for investigation of nonlinear slow optical pulse propagation in media 
with normal dispersion~\cite{Grischkowsky,Grischkowsky1}. However, a two-state system is unfit for 
observation of nonlinear superluminal effects. The problem is to observe the pulse superluminal 
propagation in a medium with an intensity-dependent refractive index, but without accompanying 
nonlinear processes, which lead to pulse reshaping as well. Since a two-level atom displays anomalous 
dispersion within the absorption line and the nonlinearity in the dispersion is achieved at the intensities 
that are much larger than the saturation intensity (see Appendix), nonlinear effects such as 
self-focusing (or defocusing), self-phase modulation etc., cannot be neglected. 
Meanwhile, the systems, that are almost transparent and at the same time show large nonlinear dispersion, 
have been basic models for theoretical and experimental studies of resonant nonlinear magneto-optical 
effects caused by a laser-induced coherence (see the recent review by Budker {\em et al.}~\cite{Budker1}). 
The simplest system displaying these properties is a three-level $\Lambda$-atom 
with ground-state Zeeman sublevels, that interacts with the right (RCP) and left (LCP) circularly 
polarized components of optical light-pulse in the presence of weak magnetic field, which is applied 
along the direction of light pulse propagation (Fig.\ref{fig1}). Due to the nearly maximal Zeeman coherence, 
large nonlinear susceptibilities are created at the resonance frequency in analogous to those observed 
in EIT experiments (for relationship between EIT and nonlinear magneto-optics see~\cite{Budker}). 
This leads to very narrow magnetic resonances~\cite{Budker2} or enhances significantly the nonlinear Faraday 
signal~\cite{Sautenkov}. However, the dynamics of resonant light propagation in coherent nonlinear media has
not been studied yet. This analysis is carried out for the first time in the present paper. For the Faraday 
configuration in Fig.\ref{fig1} we show that the light-induced Zeeman coherence preserves the anomalous 
dispersion of the medium, while it suppresses the absorption at the resonance line center. This results in
a negative nonlinear group-velocity at the values of light intensity, that are many times smaller 
than the saturation intensity of the corresponding atomic transition. In this scheme the superluminal 
effects are revealed in two ways. Firstly, similar to linear media, the peak of outgoing pulse exits the 
cell before the peak of incident pulse with a narrow spectral width enters the medium. 
Secondly, the pulse velocity is nonlinear in light intensity, the more intense parts 
of the pulse travel slower. As a result, the self-steepening of the trailing edge of pulse takes place 
in contrast to the shock-wave formation in media with normal dispersion~\cite{Grischkowsky,Grischkowsky1}.
In fact, the steepening effect is the most striking manifestation of anomalous dispersion, if one takes into 
account that usually the residual absorption causes a misconstruction of the pulse advance. 

The paper is organized as follows. In the next section we derive equations for the evolution of the atomic 
density matrix and for the field transmitted through the medium. Basing on the analytical solution for the 
density matrix we give the results of numerical calculations for the pulse propagation with nonlinear 
negative group velocity in Sec.3. Our conclusions are summarized in Sec. 4. Finally, in the Appendix we 
calculate the nonlinear group-velocity of light pulse propagating in the Doppler broadened medium for 
the cases of two-level atom and three-level $\Lambda$-system.

\section{Nonlinear negative group velocity in a medium with Zeeman coherence}

\begin{figure}[p]
\center
\includegraphics{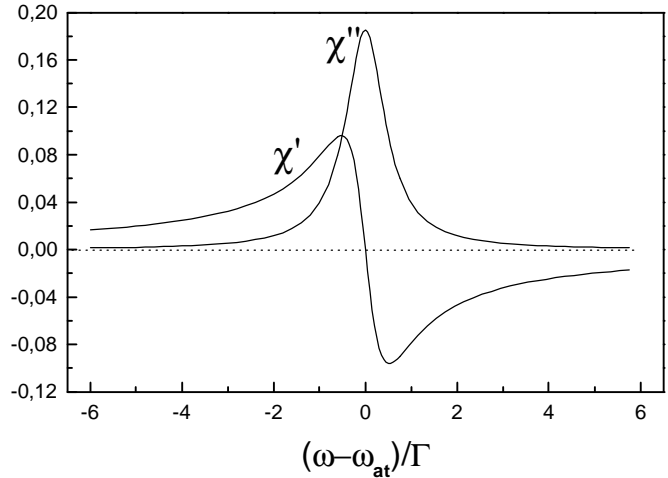}
\caption{Real $(x^{\prime})$ and imaginary $(x^{\prime \prime})$ parts of the susceptibility of two-level 
atom as a function of the atom-field detaining $\varepsilon=\omega-\omega_{at}$ in units of $\Gamma$.
\label{fig2}}
\end{figure}

\begin{figure}[p]
\center
\includegraphics{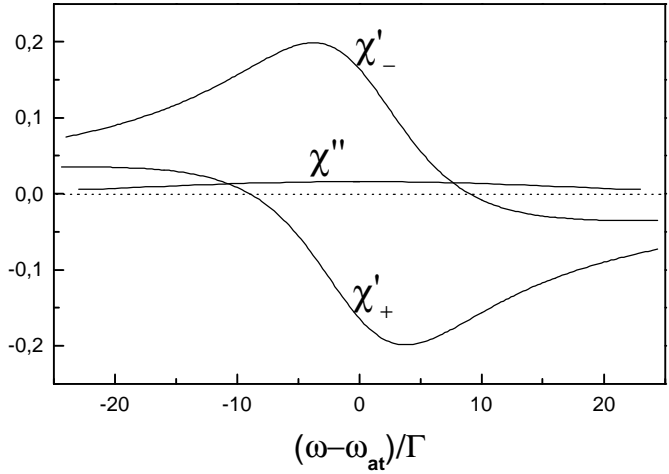}
\caption{Imaginary $(x^{\prime\prime})$ and real $(x^{\prime}_{1})$ parts of susceptibilities at 
$\sigma _{+}$ and $\sigma _{-}$ transitions in a $\Lambda$-atom as a function of 
$\varepsilon=\omega-\omega_{at}$ for $\Omega_{0}=0.3\Gamma$, $\Delta=0.01\Gamma$, $\gamma_{c}=10^{-4}\Gamma$. 
Doppler broadening is neglected.
\label{fig3}}
\end{figure}

It is well known that in linear media the anomalous dispersion always occurs within an absorption 
line~\cite{Brillouin} and leads to negative group velocity for the peak motion of sufficiently smooth 
pulses~\cite{Garrett}. However, as can be seen from Fig.\ref{fig2} for the case of a two-level atom, in 
usual materials the light pulses experience very large absorption in the vicinity of sharp atomic 
resonance that prevents a doubtless observation of high anomalous dispersion~\cite{Chiao1}. On the 
other hand, the light-induced Zeeman coherence in a three-level $\Lambda$-atom changes essentially 
this absorption-dispersion relationship, making the medium almost transparent, while keeping 
its dispersion still very large (Fig.\ref{fig3}). Moreover, as it is apparent from Fig.\ref{fig3}, there is 
a frequency range, where the dispersion on both transitions $|1\rangle \rightarrow |3\rangle$ and 
$|2\rangle \rightarrow |3\rangle$ is the same and anomalous. This means that the RCP and LCP components
of linearly polarized laser field travel with the same negative group velocity. As the latter depends on
the light intensity, the pulse advance, being the minimum at the pulse peak (see below), is different 
for different parts of the pulse compared to the same pulse traveling the same distance in a vacuum. 
As a result, the trailing edge of pulse should show self-steepening that can be tested in an 
interferometric experiment similar to that performed by L. Wang and co-workers~\cite{Dogariu}.

\subsection{Basic equations}

The atomic system depicted in Fig.\ref{fig1} interacts with the fields of RCP and LCP components $E_{1,2}$ 
of linearly polarized laser light with the frequency $\omega$ and wavenumber $k$ propagating along the $z$ 
axis. In the presence of a static longitudinal magnetic field $B$ these components,which are working on the 
transitions $|1\rangle \rightarrow |3\rangle$ and $|2\rangle\rightarrow |3\rangle$, respectively, are 
detuned from the atomic resonances by
\begin{equation}
\Delta_{1,2}=\omega-\omega_{31,2}\mp \Delta-\frac{\omega}{c}v =\varepsilon\mp \Delta,
\label{1}
\end{equation}
where $\omega_{ik}$ is the frequency difference between the levels $i$ and $k$, and $\Delta=g \mu_{B}B/\hbar$ 
is the Zeeman shift, which is induced by the magnetic field for the ground state sublevels $|1\rangle$ and 
$|2\rangle$ having magnetic quantum numbers $\mp 1$. Here $g$ is the ground-state gyromagnetic factor and 
$\mu _{B}$ is Bohr's magneton. In Eq.(\ref{1}) we have included the Doppler shift $\frac{\omega}{c}v$ 
for atoms moving with frequency $v_{z}=v$.

The interaction of fields $E_{1,2}$ with a single atom is described by the Hamiltonian:
\begin{equation}
H_{int}=\hbar \Delta _{1} \sigma _{11}+\hbar \Delta _{2} \sigma _{22}-\hbar (\Omega _{1}\sigma _{31}+%
\Omega _{2}\sigma_{32}+h.\,\,c.).
\label{2}
\end{equation}

Here $\sigma _{ij}=|i\rangle\langle j|$ are atomic operators. The Rabi frequencies of field polarization 
components are defined as $\Omega _{1,2}=\mu E_{1,2}/\hbar$, where the dipole moments of the transitions 
$|1\rangle\rightarrow |3\rangle$ and $|2\rangle\rightarrow |3\rangle$ are taken equal: 
$\mu _{13}=\mu _{23}=\mu$.

The time evolution of the system's density matrix $\rho$ obeys the master equation
\begin{equation}
\frac{d}{dt}\rho = -\frac{i}{\hbar}[H_{int}\, ,\rho]+\Lambda \rho,
\label{3}
\end{equation}
where the matrix $\Lambda\rho$ accounts for all atomic relaxations. In further, we neglect the collisional 
broadening of optical transitions compared to the spontaneous decay rate, but we take it into account for the 
ground-state coherence. This approximation is valid for the gas pressure below $10^{13}cm^{-3}$ at room 
temperature that is assumed hereafter. Thus, in our model the longitudinal and transverse optical relaxations 
of atomic Bloch vectors are determined by the process of spontaneous emission from the upper level $|3\rangle$
to the lower levels $|1\rangle$ and $|2\rangle$ with equal rates $\gamma _{1}=\gamma _{2}=\gamma$, and 
outside of the three-level $\Lambda$-system with the rate $\gamma _{0}$. The Zeeman coherence decay rate 
$\gamma_{c}$ is determined by the collisional and atomic time-of-flight broadening. Thus, the 
phenomenological relaxation matrix $\Lambda \rho$ has the form
\begin{equation}
\Lambda \rho = \left( \begin {array}{ccc}
\gamma \rho_{33} & -\gamma _{c} \, \rho _{12} & -\Gamma \rho _{13} \\
-\gamma _{c} \, \rho _{21} & \gamma \rho _{33} & -\Gamma \rho _{23} \\
-\Gamma \rho _{31} & -\Gamma \rho _{32} & -(\gamma _{0}+2 \gamma)\, \rho _{33}
\end{array} \right), 
\label{4}
\end{equation}
where $2 \Gamma =\gamma _{0}+2 \gamma$ is the total spontaneous decay rate of the level $|3\rangle$.

The evolution of the slowly varying field amplitudes $E_{1,2}$ along the $z$ axis is determined by 
Maxwell's equations
\begin{equation}
\left (\frac{\partial}{\partial z}+ \frac{1}{c}\frac{\partial}{\partial t}\right) E_{1,2}(z,t)=
2 \pi i \, \frac{\omega}{c} \, P_{1,2}(z,t),
\label{5}
\end{equation}
where $P_{1,2}(z,t)$ are the field induced polarization, real and imaginary parts of which are responsible 
for the dispersive and absorptive properties of medium at the transitions 
$|1\rangle\rightarrow |3\rangle$ and $|2\rangle\rightarrow |3\rangle$, respectively. The polarizations 
are calculated from the Fourier transform
\begin{equation}
P_{i}(z,t)=\int _{-\infty}^{\infty}d \nu \, e^{i\nu t} P_{i}(\nu , z), \quad i=1,2,
\label{6}
\end{equation}
where $\nu=\overline{\omega}-\omega$ is the deviation from the field carrier frequency $\omega$, 
and $P_{i}(\nu ,z)$ are expressed by the corresponding susceptibilities
\begin{equation}
\chi _{i} (\nu ,z)=N \mu^{2} \rho _{3i}/\hbar \Omega_{i},
\label{7}
\end{equation}
as
\begin{equation}
P_{i}(\nu ,z)=\chi _{i} (\nu ,z) E_{i} (\nu ,z), 
\label{8}
\end{equation}
with $N$ being the atomic number density.

We consider the case of smooth incident pulses with a duration much larger than all relaxation times 
of the medium. Then, the limited bandwidth of field Fourier amplitudes $E_{i}(\nu ,z)$ allows us to approximate $\chi_{i}(\nu ,z)$ 
by the first few terms of the Taylor series:
\begin{equation}
\chi (\nu ,z)=\chi(\omega ,z) + \left.\frac{\partial \chi}{\partial \nu}\right|_{\nu=0}\cdot \nu +\left. %
\frac{1}{2}\frac{\partial^{2}\chi} {\partial \nu^{2}} \right|_{\nu=0} \cdot\nu^{2}+\ldots\,\,.
\label{9}
\end{equation}
The substitution of Eq.(\ref{9}) into Eq.(\ref{6}) yields 
\begin {equation}
P_{i}(z,t)=\chi_{i}(\omega,z)\cdot E_{i}(z,t)+\left. i \frac{\partial \chi_{i}}{\partial\omega}\right|_{0}%
\cdot\frac{\partial E_{i}}{\partial t}-\left. \frac{1}{2} \frac{\partial^{2} \chi_{i}}{\partial\omega^{2}}%
\right|_{0}\cdot\frac{\partial^{2} E_{i}}{\partial t^{2}}+\ldots\,\,.
\label{10}
\end{equation}
Then, from Eq.(\ref{4}) we have
\begin{equation}
\left(\frac{\partial}{\partial z}+\frac{1}{v_{i}}\frac{\partial}{\partial t} \right)\, E_{i}(z,t)=2 \pi i \,%
\frac{\omega}{c}
\left[\chi_{i}(\omega ,z)E_{i}-\frac{1}{2}\frac{\partial^{2}\chi_{i}}{\partial \omega^{2}}\right|_{0}
\left. \frac{\partial^{2}E_{i}}{\partial t^{2}} \right],
\label{11}
\end{equation}
where group velocities $v_{i}$ ($i=1,2$) for the two polarization components are introduced in the form
\begin{equation}
v_{i}=c \left[1+\left. 2 \pi \omega\, \Re e \, \frac{\partial \chi_{i}}{\partial \omega}\right|_{0}\right]^{-1},%
\quad i=1,2.
\label{12}
\end{equation} 

This expression coincides with the classical formula of group velocity $v_{gr}=c\,\Re e\left[n+\omega\frac{d n}
{d\omega}\right]^{-1}$, if one takes into account that the complex index of refraction of the medium 
$n=1+2 \pi \chi$ and $\Re e\,\chi<<1$. The second term in rhs of Eq.(\ref{11}) is responsible for the pulse 
distortion due to the group-velocity dispersion. In the next section we will show that this term is 
negligibly small for the length of the atomic sample $L$, which is restricted by the 
requirement of smoothness of the pulse.

Since in this paper we restrict our attention to the superluminal pulse propagation, we need only the equation 
for intensities 
$I_{i}(z,t)=\frac{c}{2 \pi}|E_{i}(z,t)|^{2}$. From Eq.(\ref{11}) we have
\begin{equation}
\left(\frac{\partial}{\partial z}+\frac{1}{v_{i}}\frac{\partial}{\partial t}\right) I_{i}(z,t)=%
-\alpha _{i}I_{i}(z,t),
\label{13}
\end{equation}
where the nonlinear absorption coefficients $\alpha_{i}$ at the frequency $\omega$ are defined as
\begin{equation}
\alpha_{i}=\frac{4\,\pi\omega}{c}\,\Im m\,\chi_{i}=\frac{1}{\ell_{0}}\Im m\,\frac{\rho_{3i}\Gamma}%
{\Omega_{i}}.
\label{14}
\end{equation}
Here $\ell_{0}=(\sigma N)^{-1}$ and $\sigma =4\pi\omega\mu^{2}(\hbar c\Gamma)^{-1}$ is the resonant absorption 
cross-section.

The advance time $T_{ad}$ of the field $E_{i}(z,t)$ propagating through a medium of length $L$ compared to the 
pulse passing the same distance in a vacuum is given by
\begin{equation}
T_{ad}=\left(\frac{1}{c}-\frac{1}{v_{i}}\right) L=\left. -2 \pi \frac{\omega}{c}\, \Re e\, %
\frac{\partial \chi_{i}}{\partial \omega}\right|_{0}. \label{15}
\end{equation}
Thus, at the frequencies where the derivative $\frac {\partial \chi}{\partial \omega}$ takes a large 
negative value the pulse advance is significant.

\subsection{Solutions}

We solve Eq.(\ref{13}) assuming that the incident laser beam is a Gaussian pulse
\begin{equation}
I(z,t)=I_{in}(t-z/c)=I_{0}\exp[-(t-z/c)^{2}/\tau^{2}], \quad z \leq 0,
\label{16}
\end{equation}
with a temporal duration $\tau \gg \Gamma^{-1}$.

During the propagation of pulse through the medium the group-velocity dispersion and self-steepening 
change the pulse width that violates the approximation of truncated Taylor expansion Eq.(\ref{10}). 
Our aim is to observe the pulse nonlinear steepening under superluminal propagation without pulse spreading.
The source of the latter is the derivative 
$\partial^{2} \chi/\partial \omega^{2}$, which enters in the second term in the rhs of Eq.(\ref{11}). 
Since this term is also proportional to the second time-derivative of the field $\partial^{2} E/ \partial t^{2}$, 
the contribution from the group-velocity dispersion dramatically increases at the 
distances, where a steep front is formed due to the dependence of group velocity on light intensity. 
In order to avoid this complication, as well as to use rightly the steady-state solution of 
Eqs.(\ref{2}), we restrict the upper bound of the sample length by a value for which the pulse 
broadening caused by the self-steepening does not exceed the medium absorption line-width $\Gamma$, 
i.e. the condition 
\begin{equation}
\left| \frac{\partial I(z,t)}{\partial t}\right| \leq \Gamma I(z,t), \quad 0 \leq z \leq L,
\label{17}
\end{equation} 
should hold everywhere inside the medium for any point on the pulse envelope. Then, by the condition 
(\ref{17}), the spreading of the wave packet caused by the group-velocity dispersion can be neglected 
up to the distances restricted by the value     
\begin{equation}
L \ll L_{1}=\left| 2\,\pi\,\frac{\omega}{c} \,\Gamma ^{2} \left. \frac{\partial^{2}\chi}{\partial\omega^{2}}%
\right|_{max}\right|^{-1}.
\label{18}  
\end{equation}
Eq.(\ref{18}) follows from the smallness of the second term in the rhs of Eq.(\ref{11}). Here we
have replaced the derivative $\partial^{2}\chi/\partial\omega ^{2}$ by its maximal value taking into account 
that $\chi$ is an implicit function of $z$ through its dependence on pulse intensity.

The condition (\ref{17}) allows us to use the steady-state solution for atomic density matrix elements.
This solution is obtained from Eq.(\ref{2}) in the form
\begin{eqnarray}
\Im m\,\frac{\rho _{3i}}{\Omega _{i}}&=&\,q \Gamma \Delta^{2}\,\delta\rho/D,\quad i=1,2,\label{19} \\ 
\Re e\,\frac{\rho _{31}}{\Omega _{1}}&=&-\Delta(q\,\varepsilon\,\Delta +\Omega^{2})\,\delta\rho/D,\label{20} \\
\Re e\,\frac{\rho _{32}}{\Omega _{2}}&=&-\Delta(q\,\varepsilon\,\Delta -\Omega^{2})\,\delta\rho/D,\label{21}
\end{eqnarray}
\begin{equation}
D=q \Delta^{2}(\varepsilon^{2}+\Gamma^{2})+\Omega^{4},\quad q=1+\frac{\Omega^{2}\gamma_{c}}{2\Gamma\Delta^{2}},
\label{22}
\end{equation}
where $\Omega^{2}=|\Omega _{1}|^{2}=|\Omega_2|^{2}$. It is derived under the assumptions of 
$\Delta,\,\gamma_{c}\ll\Gamma$ and $\Gamma^{2}\geq\Omega^{2}_{0}\gg\Gamma\gamma_{c}$, where 
$\Omega _{0}$ is the Rabi frequency corresponding to the peak value of pulse intensity. The 
population difference between the ground-state Zeeman sublevels and the upper
state $|3\rangle$ is $\delta\rho$, which is $\delta\rho\simeq\frac{1}{3}$ for equal decay rates 
$\gamma_{0}=\gamma$ and a weak magnetic field. In the linear limit ($\Omega^2\ll\Delta\Gamma$) these 
expressions coincide with that for two-level atom presented graphically in Fig.\ref{fig2}. 
In the inverse case of $\Omega^2\gg\Delta\Gamma$ the medium becomes almost transparent 
near the resonance $\varepsilon=0$, while its dispersion remains highly anomalous (Fig.\ref{fig3}).
This result differs from the well-known fact established for linear media that the optical 
dispersion of transparent medium is always normal~\cite{Landau}. The latter directly follows 
from Kramers-Kronig relations, that assume linearity and causality in the response of the
medium. In our case this response is nonlinear and the classical Kramers-Kronig relations are no 
longer valid. The numerical solution of Eq.(\ref{2}) in steady-state regime is shown in 
Fig.\ref{fig3}. Our results given by Eqs.(\ref{19})-(\ref{22}) are in excellent agreement with 
these calculations.

In further, we need to take into account the Doppler broadening by averaging the matrix elements 
$\rho _{ij}$ over the atomic velocity distribution. Using the results of calculations performed in 
Appendix we finally obtain the group velocity $v_{gr}$ and the absorption coefficient $\alpha$ of 
two polarization components of the field in the form
\begin{eqnarray}
v_{gr} & = & \frac{c}{n_{g}}, \quad n_{g}=1-\frac{c \Gamma\delta\rho}{2 \ell _{0} \Delta^{2}_{D}}%
\cdot \frac{1}{(b+1)^2}, \label{23}\\
\alpha & = & \frac{1}{ \ell _{0}}\cdot\frac{\Gamma^{2}\delta\rho}{\Delta ^{2}_{D}}\cdot \frac{1}{b(b+1)}, 
\label{24}
\end{eqnarray}
where b is given by Eq.(\ref{A9}).As it follows from Eq.(\ref{23}), the group velocity index $n_{g}$ 
reaches its maximum at the peak value of light intensity. As a result, more intense parts of the 
pulse move slower that leads to the steepening of the trailing edge of the pulse.
Also, notice that in the absence of external magnetic field, when $\Delta=0$, the susceptibilities 
$\chi_i$ vanish due to complete population trapping and the pulse moves with the light vacuum velocity $c$.  

The results of numerical integration of Eq.(\ref{13}) for pulse propagation using Eqs.(\ref{23}) and 
(\ref{24}) are presented in the next section.

Now, having the solution for susceptibilities Eqs.(\ref{A6})-(\ref{A8}), we can find the upper bound of the 
cell length from the conditions (\ref{17}) and (\ref{18}). The maximal value 
$\left| \partial^{2}\chi/\partial\omega^{2}\right|_{max}=N\,\mu^{2}\delta\rho/4\hbar\Delta^{2}_{D}\Gamma$ 
occurs at $\Omega^{2}=\Delta \Gamma$. Substituting it into Eq.(\ref{18}) we have 
\begin{equation}
L_{1}=\frac{32 \pi \Delta^{2}_{D}}{3 N \lambda^{2}\Gamma\,\gamma\,\delta\rho},
\label{25}
\end{equation} 
where $\mu^{2}$ is expressed by the radiative decay rate $\gamma=32 \pi^{3}\mu^{2}/3 \hbar \lambda^{3}$
and $\lambda$ is the wavelength of atomic transition.

The condition (\ref{18}) can be rewritten in the form
\begin{equation}
\Delta \tau\ll \frac{1}{\Gamma},
\label{26}
\end{equation}
where
\begin{equation}
\Delta \tau =2\pi\omega\Gamma\left|\frac{\partial^{2}\chi}{\partial\omega^{2}}\right|\cdot\frac{L}{c},
\label{27}
\end{equation}
is the amount of spreading experienced by a pulse, when traveling through a medium with the length $L$. 
Thus, condition (\ref{26}) ensures the truly observable steepening of the backward edge of light-pulse 
without its distortion.

In order to find the upper bound for the cell length from (\ref{17}) we consider the 
situation where the absorption is absent. In this case the reduced equation (\ref{13}) has simple 
wave solution in the form
\begin{equation}
I(z,t)=I_{in}(t-z/v_{g}),
\label{28}
\end{equation}
where $I_{in}(z,t)$ is given in Eq.(\ref{16}). Since at the trailing edge $d I/d t<0$, from Eq.(\ref{27}) 
and (\ref{17}) follows
\begin{equation}
L \leq L_{2}=\frac{2\ell_{0}\tau\Delta^{2}_{D}}{\Gamma\delta\rho}\cdot\frac{(b+1)^{3}}{b}.
\label{29}
\end{equation}
 
Note that Eq.(\ref{29}) offers the most stringent restriction at the 
peak intensity where $b$ is close to the unity.
 
The last criterion is that $\Delta \tau$ should be small as compared to the group-advance 
time $T_{ad}$, $\Delta \tau \ll T_{ad}$. Using Eqs.(\ref{15}) and (\ref{27}) and solutions 
Eqs.(\ref{24}-\ref{26}) we obtain
\begin{equation}
\frac{\Gamma \Omega^{2}\delta\rho}{\Delta \Delta^{2}_{D}}\frac{2 b+1}{b^{2}(b+1)}\ll 1 \label{30}
\end{equation}

One can easily show that this condition fails only for sufficiently small 
$\Omega\sim \sqrt{\Delta \Gamma}$, so that there is no reason to consider this complication here.

It must be also noted that for the group-advance to be observable, it is sufficient that $T_{ad}$ 
is large with respect to the resolution time of interferometric detection, being at the same time much 
smaller than the pulse width $\tau^{-1}$, as it was in the recent experiment~\cite{Dogariu}. 

\section{Numerical results and discussion}

Here we give results of numerical calculations based on the equations derived in the preceding section and 
discuss the relevant physics.

In our simulations for the model of atomic system we have chosen parameters corresponding to the $D_{1}$ 
absorption line of $^{87}Rb$. It is worth noting that a four-state system has been employed in 
~\cite{Sautenkov} for investigation of nonlinear magneto-optical effects in rubidium.It has been shown that 
this simplified model, with well-chosen parameters, represents the qualitative physics quite 
well~\cite{Sautenkov}. In our case the total decay rate of the upper level $|3\rangle$ is chosen 
$2\Gamma =2\pi\times5\cdot 10^{6} \,s^{-1}$, the wavelength of optical transitions $\lambda =800\, nm$,
and the inhomogeneous Doppler broadening width $\Delta _{D}\simeq 50\,\Gamma$. For the atomic number 
density $N=2 \cdot 10^{12}\, cm^{-3}$ we have $\ell _{0}\simeq 1.25 \cdot 10^{-4}\, cm$. The external 
magnetic field is taken $B=30\,mG$, which corresponds to $\Delta =0.002\, \Gamma$ for the ground-state level
of rubidium. The peak value of Rabi frequency of the field is chosen to obey the condition 
$\Omega^{2}_{0}=3\,\Delta \Delta _{D}$, i.e. $\Omega _{0}=0.6 \, \Gamma$ ($I_{0}=5\,mW/cm^{2}$), which is below 
the saturating value for a two-level atom $\Omega _{s}=2 \,\Gamma$. The Zeeman coherence relaxation rate 
$\gamma _{c}$ is attributed to the collisional (in the presence of buffer gas) and time-of-flight broadening 
and is taken $\gamma _{c}\leq 10^{-4}\, \Gamma$. Then, for initial pulse duration $\tau=3\,\mu s$ and 
population difference $\Delta \rho \simeq 0.25$ the upper bounds of the sample length are estimated from 
Eqs.(\ref{25}) and (\ref{29}) as $L_{1}\simeq 4 \cdot 10^{4}\, cm$ and $L_{2}\simeq 400\, cm$, 
respectively. This means that in the cell with a length of a few centimeters the superluminal effects 
including the pulse self-steepening can be easily observed with negligible pulse distortion.

In Fig.4 we plot the absorption of the pulse and the group-advance time calculated from Eqs.(\ref{24}) 
and (\ref{15}) as a function of the pulse intensity for the set of parameters given above. It is 
apparent that the pulse peak is advanced with respect to the same pulse traveling in the vacuum 
approximately fifty times, that for $L=6\,cm$ corresponds to $T_{ad}\simeq 10\,ns$. The advance time
increases towards to the pulse wings, and it is about $70\,ns$ for the intensity $I\sim 0.1\,I_{0}$. 
Since the absorption is strongly 
reduced around the pulse peak, where the EIT conditions 
\begin{equation} 
\Omega^{2}_{0}\gg \Delta\Delta_{D},\gamma_{c}(\Delta_{D}+\Gamma),
\label{31}
\end{equation}
are established, the group-advance time should be easily observed by nanosecond resolution time.

As an illustration, in Fig.\ref{fig5} we give a space-time plot of solution to Eq.(\ref{13}) for the initial
Gaussian pulse with duration $\tau\simeq 1\,\Gamma^{-1}$. The group-velocity is negative in the medium. 
This effect appears as a "back-on-time" motion of the pulse. The formation of sharp back front of the pulse 
to the end of the cell is well visible (Fig.\ref{fig5}).

In order to reconstruct the superluminal propagation in details we present in Fig.\ref{fig6} the pulse 
motion in the form of time-sequence graphs. At first, before the incident pulse enters the medium, 
a weak pulse is created at the output. Then, this peak increases in time and eventually splits into 
forward- and backward-moving pulses with very sharp trailing edges. The first of them continues to move 
in a vacuum with velocity $c$, and after a long time it represents the final pulse. The second pulse exhibits, 
however, an unusual behavior. It increases while traveling through the absorbing medium. Nevertheless, 
here there is no inconsistency with the energy conservation law. Firstly, the velocity of energy 
transport defined as $v_{E}=\overline{S}/\overline{W}$, where $\overline{S}$ and $\overline{W}$ are 
the energy flux and field energy densities averaged over a period, is always less than $c$, as is shown 
in~\cite{Diener2}. Secondly, the increment of the medium energy in the presence of light field is proportional 
to $\partial(\omega n)/\partial \omega \cdot |E|^{2}$~\cite{Landau} and is negative for 
anomalous-dispersive media. Obviously this energy cannot be  stored in the medium and it is transformed 
into the backward-moving wave. However, this takes place until the peak of incident pulse enters the medium. 
After that, the real absorption of the light pulse results in a cancellation of backward wave at
the entrance of the cell.

In Eq.(\ref{13}) the nonlinear absorption and intensity-dependent group-velocity are responsible for
the reshaping of pulse. In order to demonstrate the influence of the two mechanisms separately we solved
Eq.(\ref{13}) in a hypothetical case of nonlinear absorption, assuming that the group-velocity is independent
on the intensity of light. Setting b=0 in Eq.(\ref{23}), we calculated a propagation of Gaussian pulse
with the same set of parameters. The results depicted in Fig.\ref{fig7} show that in this case the final 
pulse is built up much earlier than in the case of nonlinear group-velocity and, what is most important, 
it preserves the shape and width of incident pulse. 
A slight decrease of absorption is observed also. 

\section{Summary}

In this paper we analyzed the superluminal propagation of light-pulse in a nonlinear medium and showed that 
necessary conditions for these effects to be observable are realized in a three-level $\Lambda$-system 
interacting with a linearly polarized laser beam in the presence of a static magnetic field. It is highly 
important that the nonlinearity of refractive index of the medium arises in low power regime, when 
all other nonlinear processes are negligible. Our results demonstrate that the light-induced Zeeman 
coherence cancels the resonant absorption of the medium almost completely, but it preserves the 
dispersion anomalous and very high. As a result, the transparent propagation of light pulse with a 
negative group-velocity, which is nonlinear in the light intensity, is predicted. This leads 
to the formation of extremely sharp trailing edge and to the lengthening of the leading edge of the pulse.
Such a behavior is inverse to the case of nonlinear propagation of the pulse in a medium with normal 
dispersion. The predicted effect is the most striking manifestation of superluminality and it can 
be easily observed in the well known schemes that have been used for studying the nonlinear magneto-optical 
rotation. 

The discussion of many questions remained behind the scope of this paper. In particular, we have not given 
due attention to the proof of the fact that the predicted effect does not violate the causality. Here we 
note only that the proof is carried out in the same manner as in previous studies of superluminal 
propagation~\cite{Diener}. Further, for real atomic systems the complete energy state description of the multilevel structure should be included. This question requires a careful examination and it is in current study, 
the results of which will be presented in the next publication.

\section*{Acknowledgments}

This work was supported by NFSAT-CRDF Grant No PH-071-02 (12015) and in part by the Government of the Republic 
of Armenia (Scientific Project No.1304).

\appendix
\section*{Appendix}
\setcounter{equation}{0}
\renewcommand{\theequation}
{\mbox{A.\arabic{equation}}}

Here we calculate the nonlinear group-velocity of light pulse propagating through a Doppler broadened 
atomic medium. The cases of two-level atom and three-state $\Lambda$-system are considered. In order to obtain 
the simple analytical expressions, when averaging the group velocities over the atomic thermal motion, we 
approximate the usual Gaussian distribution of atomic velocity with Lorentzian function   
$W(v)=\Delta_D /\pi [\Delta_D^2+(kv)^2]$, where $2\Delta_D$ is full width half maximum of
Doppler broadening.

In the case of two-level atom the imaginary and real parts of the susceptibility of medium are given by
\begin{eqnarray}
\Im m\, \chi &=& N \mu^2\Gamma / \hbar\left( \Gamma^2+\varepsilon^2+2\Omega^2\right), \\
\Re e\,\chi &=& N \mu^2\varepsilon / \hbar\left( \Gamma^2+\varepsilon^2+2\Omega^2\right),
\end{eqnarray}
where $\varepsilon=\omega_{at}-\omega+k\,v$, and $\Omega$ is the Rabi frequency of the field, which couples the 
upper and lower levels of the atom.  Straightforward calculation of the velocity-averaged susceptibilities 
in the first order of $\Delta \omega=\omega-\omega_{at}$ gives
\begin{eqnarray}
\left\langle \Im m\,\chi\right\rangle &=&\frac{N\mu^2\Gamma}{\hbar b_0(b_0+\Delta_D)}, \label{A3} \\
\left\langle \Re e\,\chi\right\rangle&=&-\frac{N\mu^2}{\hbar (b_0+\Delta_D)^2}\Delta\omega, \label{A4}
\end{eqnarray}
where $b_0=(\Gamma^2+2\Omega^2)^\frac12$.For the group-velocity of light pulse  from Eq.(\ref{A4}) we find 
\begin{equation}
v_{gr}=c\left(1+2\pi\omega\,\frac{\partial\left\langle\Re e\,\chi\right\rangle}{\partial\omega}\right)^{-1}%
\approx-\frac{c\hbar(b_0+\Delta_D)^2}{N\omega\mu^2}.
\label{A5}
\end{equation}
Similarly, for the case of three-level $\Lambda$-atom, from Eqs.(\ref{19})-(\ref{22}) we have 
\begin{eqnarray}
\left\langle \Im m\,\chi_{i}\right\rangle&=&\frac{N\mu^{2}\Gamma}{\hbar\Delta_{D}^{2}}\cdot%
\frac{\delta\rho}{b(b+1)}\cdot\left[ 1-\frac{2 b+1}{4 \Delta^{2}_{D}b^{2}(b+1)}\Delta\omega^{2}+\ldots\right],%
\label{A6}\\
\left\langle \Re e\,\chi_{1}\right\rangle&=&-\frac{N\mu^{2}}{\hbar\Delta_{D}^{2}}\cdot\frac{\delta\rho}{b+1}%
\cdot\left[\frac{\Omega^{2}}{q\,b\,\Delta}+\frac{\Delta\omega}{b+1}-\frac{\Omega^{2}}{2\,q\,\Delta%
\Delta^{2}_{D}}\cdot\frac{2b+1}{b^{2}(b+1)^{2}}\cdot \Delta \omega^{2}+\ldots \right], \label{A7}\\
\left\langle \Re e\,\chi_{2}\right\rangle &=& \Re e\chi_{1}(\Omega^{2}\rightarrow -\Omega^{2}), \label{A8}
\end{eqnarray} 
where
\begin{equation}
b=\frac{\sqrt{\Delta^{2}\Gamma^{2}+\Omega^{4}q^{-1}}}{\Delta \Delta _{D}}.
\label{A9}
\end{equation}

Correspondingly, for the group velocity of polarization components of the field is obtained in the form of
Eq.(\ref{23}). The comparison of Eqs.(\ref{A5}) and (\ref{23}) shows that, for the case of two-level atom, 
the dependence of $v_{gr}$ on light intensity appears at $\Omega \geq \Delta_D \gg \Gamma$, whereas in a 
three-level $\Lambda$-system it is achieved at much smaller values 
of $\Omega\geq (\Delta\Delta_D)^\frac{1}{2}$.

\begin{figure}[p]
\center
\includegraphics{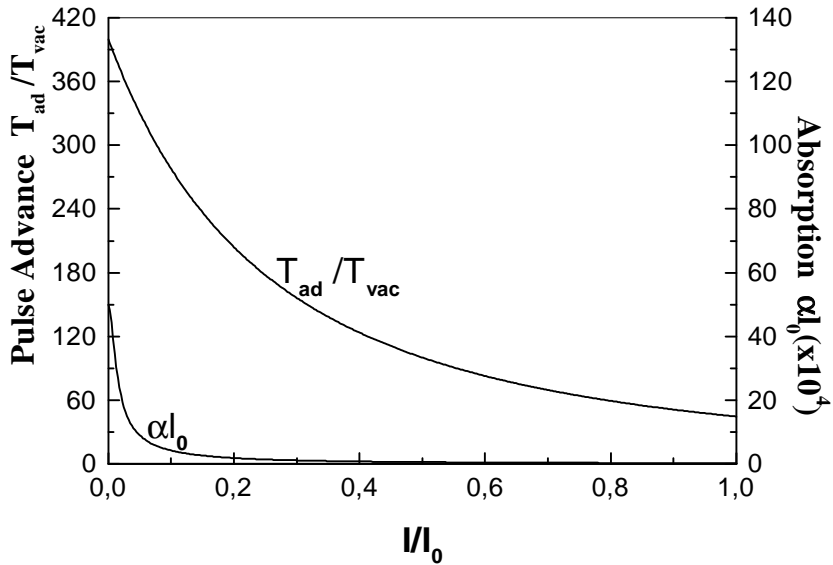}
\caption{Absorption $\alpha$ (in units of $\ell^{-1}_{0}$) and group-advance time $T_{at}$ (in units of 
$L/c$) as a function of 
light intensity. The parameters are given in the text.
\label{fig4}}
\end{figure}

\begin{figure}[p]
\center
\includegraphics{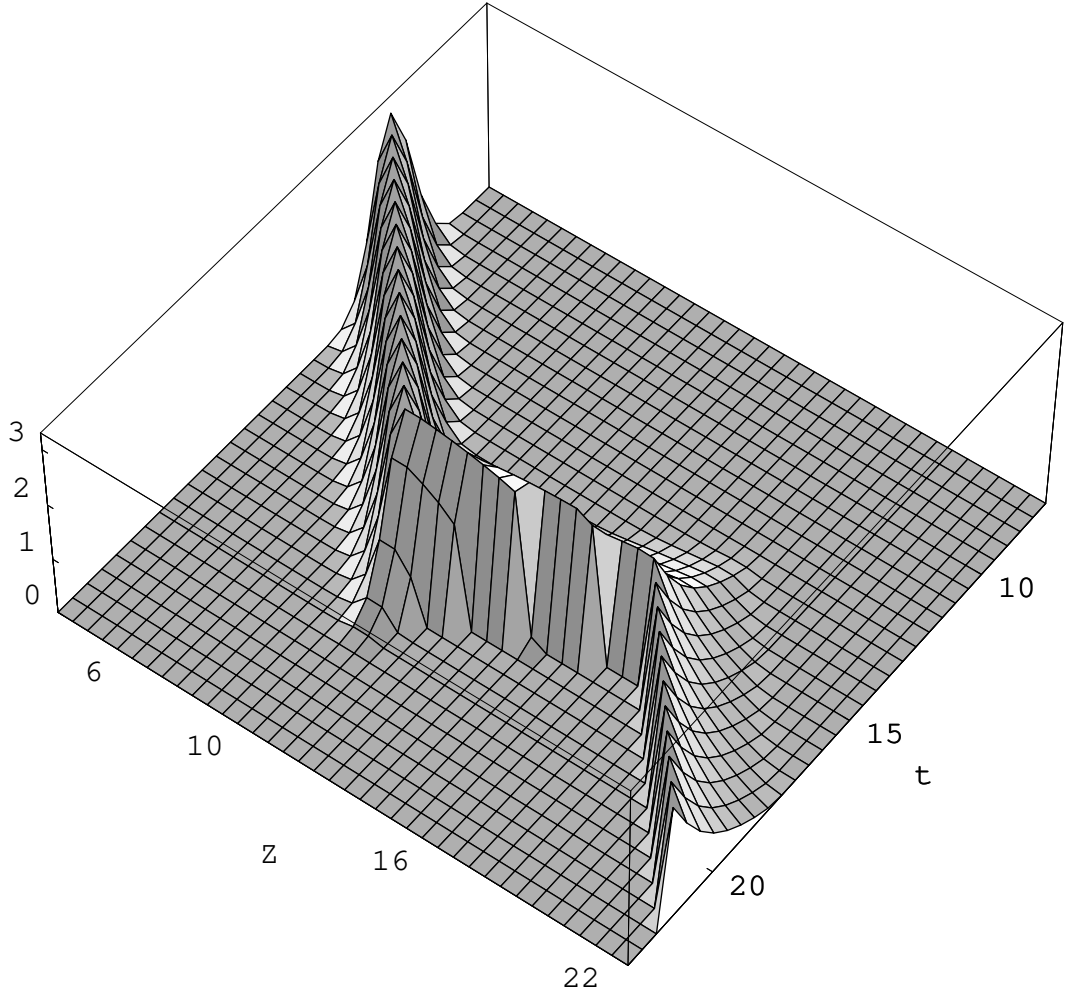}
\caption{Space-time image of Gaussian pulse motion with initial pulse
duration $\tau=1\Gamma^{-1}$. Times are given in units  of $ \Gamma^{-1}$ and
distances in $ c\Gamma^{-1}$.  The medium extends from $z=10$ to $16$. The
peak of the pulse  is at $z=-7$ at $t=0$. The group velocity and absorption coefficient 
are calculated for the parameters given in the text.
\label{fig5}} 
\end{figure}

\begin{figure}[p]
\center
\includegraphics{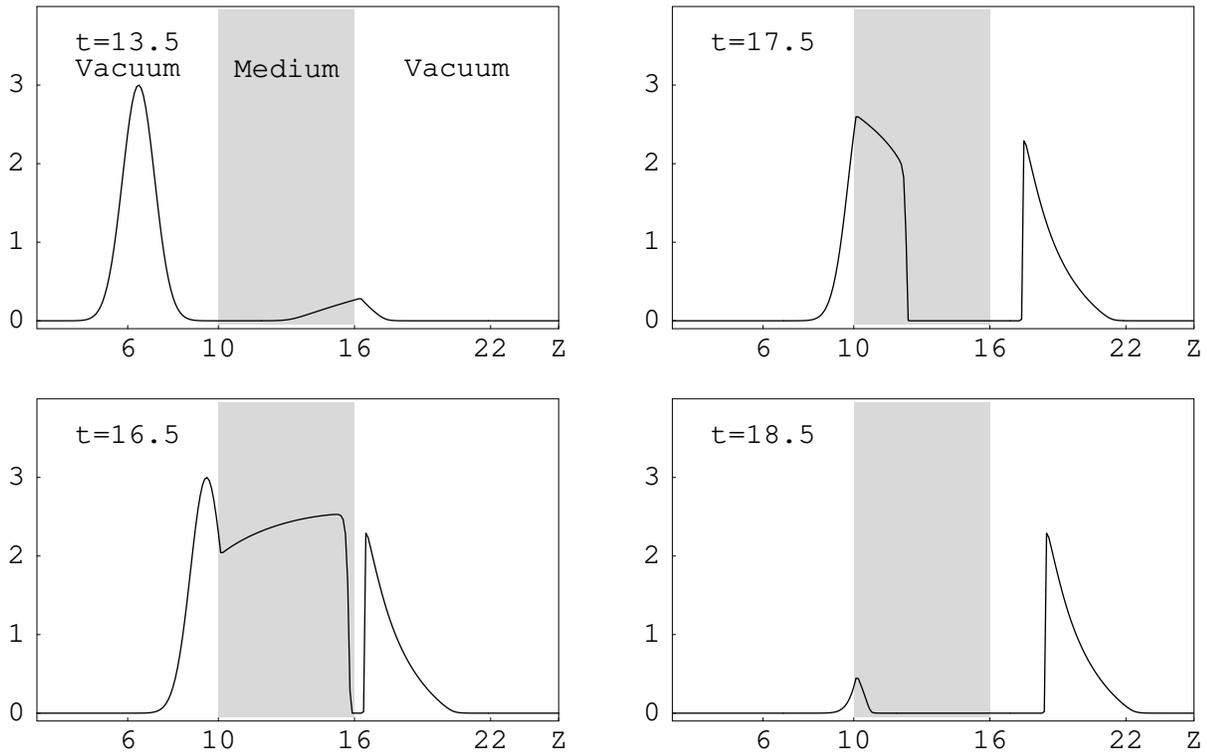}
\caption{Diagrams of Gaussian pulse propagation corresponding to sequential
instants in Fig.5.
\label{fig6}}
\end{figure}

\begin{figure}[p]
\center
\includegraphics{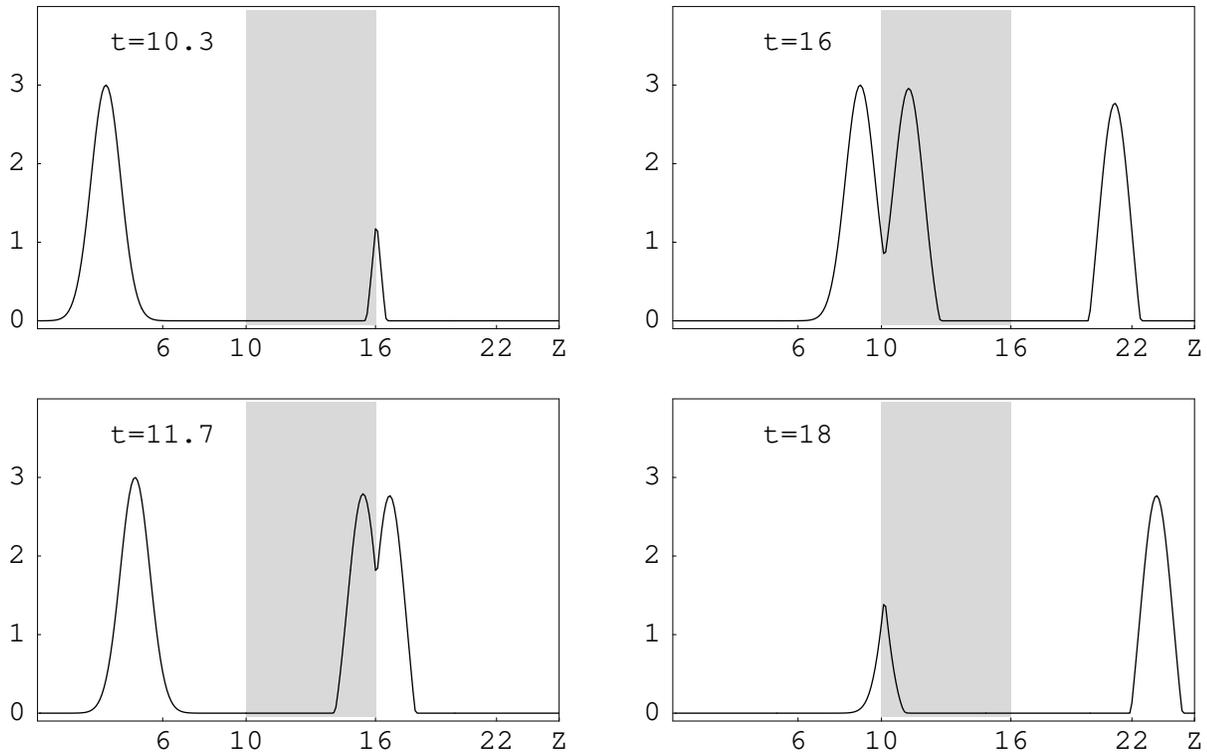}
\caption{Same configurations as for Fig.6, but in the hypothetical case of 
nonlinear absorption without intensity dependence of group-velocity.
\label{fig7}}
\end{figure}

\end{document}